# *In situ* Biological Particle Analyzer based on Digital Inline Holography

Delaney Sanborn [1,2], Ruichen He [1,2], Lei Feng [2], Jiarong Hong [1,2]

*1. Mechanical Engineering, University of Minnesota, Minneapolis, MN 55455, USA*
*2. Saint Anthony Falls Laboratory, University of Minnesota, Minneapolis, MN 55414, USA*

Note: Delaney Sanborn and Ruichen He contributed equally to this work.

Corresponding author: Jiarong Hong (jhong@umn.edu)

## *Abstract (200)*

Obtaining *in situ* measurements of biological microparticles is crucial for both scientific research and numerous industrial applications (e.g., early detection of harmful algal blooms, monitoring yeast during fermentation). However, existing methods are limited to offer timely diagnostics of these particles with sufficient accuracy and information. Here, we introduce a novel method for real-time, *in situ* analysis using machine learning assisted digital inline holography (DIH). Our machine learning model uses a customized YOLO v5 architecture specialized for the detection and classification of small biological particles. We demonstrate the effectiveness of our method in the analysis of 10 plankton species with equivalent high accuracy and significantly reduced processing time compared to previous methods. We also applied our method to differentiate yeast cells under four metabolic states and from two strains. Our results show that the proposed method can accurately detect and differentiate cellular and subcellular features related to metabolic states and strains. This study demonstrates the potential of machine learning driven DIH approach as a sensitive and versatile diagnostic tool for real-time, *in situ* analysis of both biotic and abiotic particles. This method can be readily deployed in a distributive manner for scientific research and manufacturing on an industrial scale.

## I. INTRODUCTION

Microparticles are ubiquitous in nature (dust, liquid droplets, sand, spores, fungi, bacteria etc.), and commonly appear in many industrial applications (e.g., manufacturing, food, cosmetics, pharmaceutical). Particularly, biological particles, which are derived from biological organisms including bacteria, fungi, algae, and cells, play important roles in the environment, human health, and industrial production. Technologies that can accurately characterize these particles (concentration, size, shape, composition, viability, etc.) *in situ* – that is, in their natural environment or during industrial processes – in a timely fashion are critical. These technologies have numerous applications. For example, in medical diagnosis, the ability to analyze blood cells and detect abnormalities such as circulating tumor cells (CTCs) can help determine the course of certain cancers and corresponding treatments (Yu et al., 2014). In the production of alcoholic beverages, such as beer and wine, the concentration, viability, and vitality of yeast cells need to be closely monitored for fermentation control to achieve the desired taste profile of the final product (Mochaba et al.,1998; Heggart et al., 2000). It is also important to detect and identify any type of wild yeast contamination during the process to avoid spoilage. In the field of environmental science, long-term *in situ* monitoring of different algal species and their concentrations in aquatic environments can help researchers understand the causes and dynamics of harmful algal blooms (HABs) and facilitate early detection and proper mitigation strategies to reduce their detrimental environmental and economic impacts (Ho & Michalak, 2015).

Currently, the commonly used tools for biological particle analysis are based on light or acoustic scattering or coulter principle (Maltsev & Semyanov, 2013; Baddour et al., 2006; Sun & Morgan, 2010). Light and acoustic scattering methods capture forward, side, or backward scattered signals from the particles in the sample volume and derive their concentration and size distribution based on the assumption of their shapes and scattering properties. The coulter method detects the momentary changes in impedance as a voltage pulse when suspended particles go through the orifice in the electrolyte solution. The pulses can be used to measure particle counts and volume (size) represented in terms of equivalent spherical diameter. Although these tools provide rapid measurements of particle counts and size distribution with high throughputs, they lack fidelity for the characterization of biological particles due to the complex scattering properties associated with their shapes (many non-spherical) and non-uniform internal structures. As a result, these methods often cannot distinguish different types of particles (e.g., fungi, mold, strains of bacteria), nor can they provide additional important information, such as morphology or other physiological characteristics (e.g., viability or vitality). Autofluorescence spectroscopy is a recently developed technique for particle analysis that utilizes laser-induced fluorescence (LIF) to detect molecules that absorb laser light and emit it at a higher wavelength (Croce & Bottiroli, 2014). While LIF can differentiate between biological and inert particles, it is not able to identify specific types of microorganisms or distinguish between live and dead cells, which is important for many industrial applications. Other methods, such as combining light scattering with flow cytometry, provide additional viability counting (Davey & Kell, 1996; Shapiro 2004). However, these methods also cannot classify different types of particles and require special reagents to stain cells, limiting their use in the *in situ* monitoring, especially in natural environments.

In contrast, many laboratory-based particle analysis tools can detect and identify individual biological particles by obtaining additional information. For example, microscopic imaging is a commonly used method for obtaining morphological information. In conjunction with fluorescent staining, it can differentiate biological particles with similar morphological features with high specificity and determine viability (Coling & Kachar, 1998; Stephens & Allan, 2003). However, these methods are labor-intensive and low throughput, requiring complicated sample preparation, which limits their use for *in situ* measurements or inline monitoring. To address these limitations, several label-free methods have been developed for biological particle analysis in recent years. Raman spectroscopy, for example, has been used to rapidly identify pathogen bacteria using the unique molecular compositions that result in subtle differences in their corresponding Raman spectra (Strola et al., 2014; Ho et al., 2019). However, this method requires samples with a high concentration of pure cells and cannot distinguish different organisms in mixed samples. Hyperspectral microscopic imaging can classify single cells of foodborne pathogens (Yoon et al., 2009; Eady et al., 2015; Kang et al., 2020), while quantitative phase imaging (QPI) has been used to extract detailed information about the biochemical composition of various biological particles (Popescu 2011), such as change of polyhydroxyalkanoates (Choi et al., 2021) and chromosomes (Sung et al., 2012) in individual live bacterial cells. These methods, however, require complicated optical setups and computationally intensive postprocessing, making them difficult to use for *in situ* particle analysis.

Since the beginning of the 21st century, digital inline holography (DIH) has emerged as a compact, label-free approach for the *in situ* characterization of particles (Katz & Sheng, 2010; Kaikkonen et al., 2014; Nayak et al., 2019; Sauvageat et al., 2020). This method utilizes a coherent light source, such as laser, to illuminate a three-dimensional (3D) sample volume. A digital sensor, such as a camera, records (without focusing) the interference pattern generated by the scattered light from individual particles and non-scattered portion of the illumination beam (referred to as holograms). The recorded hologram contains the phase and intensity information of the sample, which can be used to derive the 3D location, size, morphology, and optical density of particles through digital reconstruction using different diffraction formulas (e.g., Fresnel and Rayleigh-Somerfield). In comparison to conventional microscopic imaging, DIH offers orders of magnitude larger depth of field and richer information about particle properties, as the optical properties of the particles can potentially be correlated with their biochemical compositions (Beuthan et al., 1996; Choi et al., 2010; Bista et al., 2011). However, conventional DIH has several issues related to data processing, such as high computational cost and low signal-to-noise ratio due to noises from cross interference between particle signals, which limits its widespread adoption as an *in situ* tool.

To address the challenges of DIH data processing, machine learning (ML) has been recently introduced. For example, Shao et al. (2020a, b) proposed a modified U-net architecture for the fast extraction of 3D particle positions and size distribution directly from the holograms without conventional reconstruction steps. Other studies have implemented ML models for classifying cancer cells with molecule specific microbeads attached (Kim et al., 2018) and different species of plankton (Guo et al., 2021) from raw holograms. While these machine learning approaches have improved processing speed, they still require preprocessing steps such as object detection and segmentation before classification, adding complexity and computational burden. As a result, the implementation of DIH for real-time *in situ* data processing remains a challenge.

To address this challenge, we introduce a real-time hologram analysis approach based on the powerful one-stage detection and classification machine learning architecture, You Only Look Once (YOLO). YOLO has been widely used in computer vision with fast processing speed while maintaining high accuracy (> 80%) (Redmon et al., 2016). The YOLO architecture performs both object localization and object classification simultaneously, making it exceed the traditional deformable part model (DPM) and two-stage CNN based methods like R-CNN in object detection tasks with a 10x faster processing speed (Yan et al., 2014; Glenn et al., 2021). However, YOLO models are typically designed for conventional photographic objects (e.g., animals, license plate, plant) which are always in focus or close to focus and are composed of many well-defined features like contours and texture. In contrast, objects in holograms produce diffraction patterns that change significantly based on their 3D positions. It remains to be investigated whether and how the YOLO model can be adapted for hologram processing to develop accurate, robust, and real-time DIH for *in situ* biological particle analysis.

This paper is organized as follows. In the Materials and Methods section, we introduced and described a customized YOLO model for detecting and classifying individual biological particles from enhanced holograms without any additional steps. In the Results section, we demonstrated the effectiveness of our approach by applying it to classify 10 different species of plankton and differentiate yeast cells under different metabolic states and different strains during fermentation.

Finally, in the Conclusions and Discussions section, we summarized our findings and discussed their implications.

## II. MATERIALS and METHODS

The DIH setup used in this study is illustrated in Fig. 1a. It uses a collimated laser as the illumination light source. The beam passes through the sample volume and the digital camera with a microscopic objective captures the fringe patterns (holograms) generated by the interference between the scattered light from the sample and non-scattered portion of the illumination beam. The recorded holograms are then passed to the processing board (e.g., CPU or GPU) for analysis. Raw holograms are enhanced through a moving window background subtraction to eliminate the noise associated with slow variation in background over time due to, for example, the change in light source intensity during the recording. The background for each hologram is calculated as the average intensity of the 50 consecutive holograms immediately prior to it. The enhanced holograms are obtained by subtracting the background from the raw hologram.

A modified YOLOv5 machine learning architecture was used to detect and classify different biological particles in the enhanced holograms. The proposed YOLO architecture consists of three components (Fig. 1b): a backbone that extracts a collection of features (e.g., edges and corners of objects) from the input holograms, a neck that combines feature maps from different scales to generate a feature pyramid, and a prediction head that localizes and classifies individual objects based on the feature pyramid. The backbone is comprised of five convolution layers with a 3x3 kernel size. A shallower layer is used to accommodate the smaller size of the biological particles in our application. Small objects occupy few pixels, and the key features are progressively lost when passing through many convolutional or pooling layers (Nguyen et al., 2020). Reducing the number of convolutional layers in the backbone can also increase the efficiency of the model while maintaining the same level of accuracy. A maximum pooling layer is used after each convolution layer to reduce the dimensions of the feature map by summarizing the features presented in a local region. The pooling layer in our model reduces computational cost by decreasing the input image size by half, while also improving robustness to input variance using a filtered feature representation that helps prevent overfitting. The extracted feature map is then passed through three additional convolution layers in the neck with kernel size of 3x3, 3x3, and 1x1 respectively. The output of the downscaled feature map is fed into the prediction head through two separate pathways. In one of these pathways, an up-sampling layer is employed to generate the upscaled feature map, which is then combined with the original down-sampled feature map from the backbone via a concatenate layer. At the end of each pathway, a fully connected layer serves as the prediction head, which predicts the location of the objects, the confidence score associated with each object, and the class probabilities. Leaky ReLU activation function (Maas et al., 2013) is utilized after each convolution layer in the YOLO architecture, and stochastic gradient descent (SGD) is used for optimization during the training of the model. The Binary Cross Entropy with Logits Loss function is used for the loss calculation of class probability and object score. A non-maximum suppression step (Neubeck & Van Gool, 2006) is applied as a post-processing step to refine the initial prediction result and provide the final location and classification of each object in the input holograms. The proposed machine learning method is implemented using PyTorch and optimized using TensorRT. All test cases were run on a Nvidia V100 Tensor Core GPU.

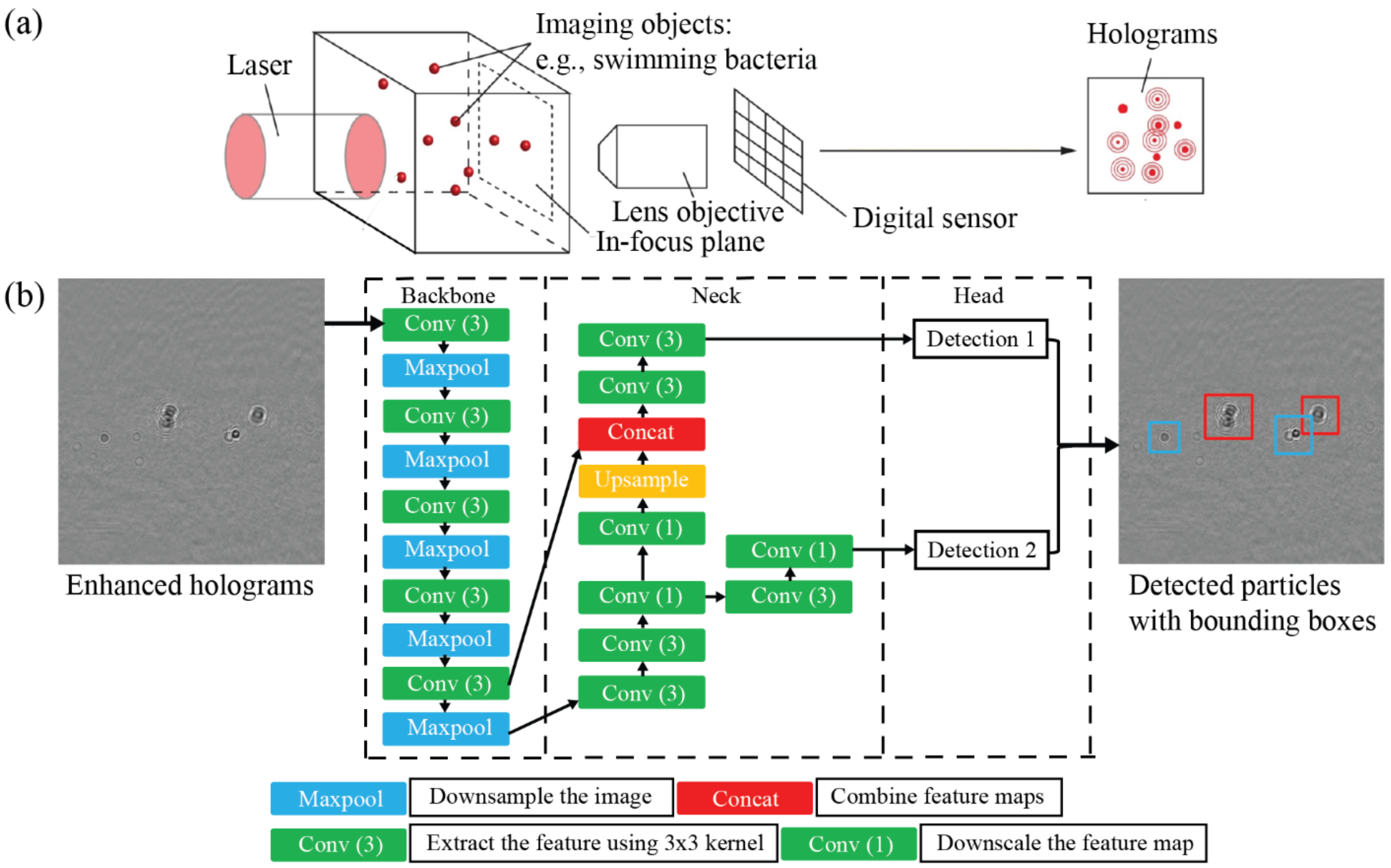

**FIG. 1.** (a) Schematic showing DIH setup for imaging particles in a 3D volume. (b) Schematic illustration of the proposed machine learning method for particle detection from holograms. Note that the input of the method is the hologram without reconstruction. The architecture of the customized YOLOv5 is described by the flowchart in the boxes with dashed lines.

In this study, we used three datasets to evaluate the performance of our method. These datasets consist of 10 different plankton species, ale yeast under four different metabolic states during fermentation, and two strains of yeasts (ale and lager) under the same fermentation condition. The plankton dataset was obtained from Guo et al. (2021) and included holograms of 10 plankton species captured by a submersible digital holographic imaging system (HOLOCAM) at two different locations. The plankton species included in the dataset are *Chaetoceros debilis, Diatom sp., Ditylum brightwelli, Chaetoceros concavicornis, Thalassiosira sp., Copepod, Copepod nauplius, cf. Strombidium sp., Tripos cf. muelleri, and Tripos cf. furca* (classified as type 0 to 9, respectively). The HOLOCAM used a 660 nm pulsed Nd-YAG laser as the illumination source and a 2048 × 2048 pixel CCD camera to record the holograms. Fig. 5a shows sample holograms and their corresponding in-focus images. The plankton species captured in the holograms are typically on the scale of several hundreds of microns, and the field of view (FOV) of the images is 9.4 x 9.4 mm with a resolution of 4.59 μm per pixel, which is sufficient to resolve unique features of the plankton species for classification (e.g., size, shape, diffraction pattern). To create the dataset, single cells were cropped from the enhanced holograms in the study by Guo et al. (2021) and divided into training database (70%), validation database (20%) and test database (10%). Table 1 summarizes the number of single cells of each sampled plankton species in the training and test datasets. The validation cell numbers for each species are included in the training dataset summary. Differences in the number of plankton cells represented in each class reflect the natural distribution of different species in the sampled water body during the HOLOCAM deployments. Additional details about the plankton dataset can be found in Guo et al. (2021). Synthetic holograms were generated by combining multiple randomly selected single-particle holograms from the database created by Guo et al. (2021) to create new training, validation, and test datasets. These datasets were used to train a YOLO model that can detect different types of plankton species.

**Table 1.** Summary of the number of plankton cells used for training and testing.

| | 01 | 02 | 03 | 04 | 05 | 06 | 07 | 08 | 09 | 10 |
|---|---|---|---|---|---|---|---|---|---|---|
| Train | 2754 | 7299 | 15120 | 5832 | 4770 | 3339 | 1854 | 5589 | 8208 | 21150 |
| Test | 306 | 811 | 1680 | 648 | 530 | 371 | 206 | 621 | 912 | 2350 |

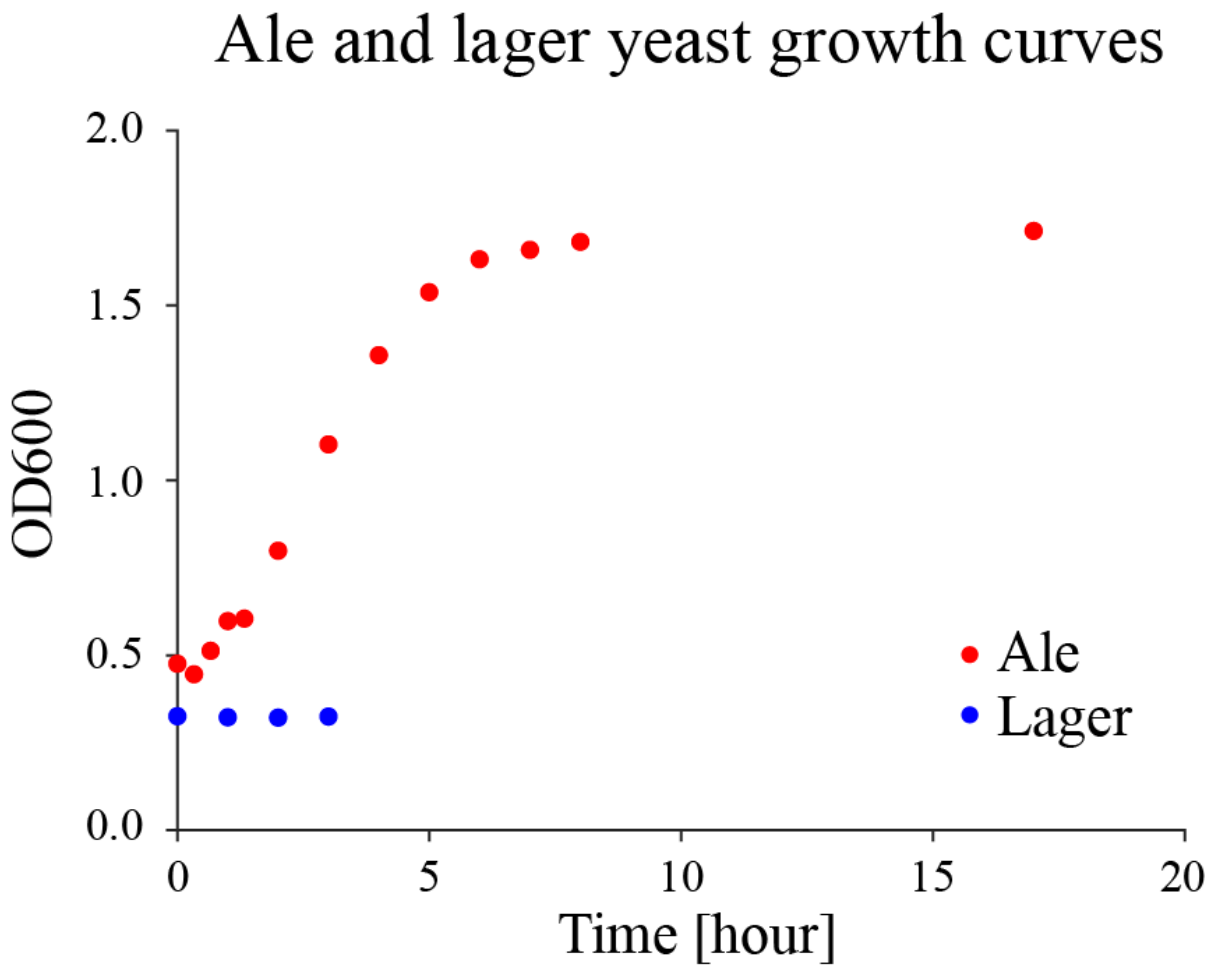

**FIG. 2.** Growth curve of the ale yeast (red dot), and lager yeast (blue dot) strain fermented under 30 ºC conditions. The cell concentration is estimated as the optical density (OD) measured at a wavelength of 600 nm.

The yeast dataset was generated by culturing two different strains of yeast cells, *Saccharomyces cerevisiae* (ale yeast, Safale US-05) and *Saccharomyces pastorianus* (lager yeast, Lallemand Diamond), under certain fermentation conditions. For metabolic state analysis, dry ale yeast was dissolved in sterile YPD media (Sigma Aldrich) and cultured overnight at 30 ºC for 16 hours. The overnight culture was then centrifuged to remove the liquid and diluted into fresh YPD to an initial optical density of approximately 0.5 at 600 nm. The diluted culture was divided into 20 milliliters samples which were used to capture holograms at 0, 1 and 4 hours, corresponding to the start, lag (cells are adapting to their new environment and division has not yet begun), and log (cells start to divide and cell number rapidly increases) phases on the growth curve (Fig. 2). An extra 20 milliliter sample was imaged at 54 hours as the dead group, making a total of four groups. The holograms were captured using a FLIR camera (Blackfly S USB3) with a purple laser diode (405 nm wavelength) and a 10X objective lens. The size of the captured holograms is 1440 x 1080 pixel with a resolution of 0.34 μm/pixel. According to the Nyquist sampling theorem (Nyquist, 1928), this resolution can resolve features in the images that are larger than 0.68 μm, which is sufficient to distinguish the cellular features of yeast cells, typically 5-10 μm in diameter as well as well as subcellular features such as vacuoles that are 1-5 μm in size. During imaging, the yeast sample flowed through a customized microfluidic channel with sheath and main flows, with a flow rate of 8 μL/min for the sheath (distilled water) and 0.4 μL/min for the main flow (yeast sample). The width of the imaging channel was 1 mm and the depth was 0.5 mm. The fringe patterns (i.e., the spacing and width of fringes) change with the distance between yeast cells and the hologram recording plane. If the focal plane of the cells is in the center of the 0.5 mm deep channel, the cells could appear anywhere from 0 to about 250 μm from the focal plane (given varying focal distances). To train a machine learning model which can detect and characterize cells regardless of the size and fringe pattern differences associated with the recording depth, we included consistent numbers of cell images varying near and far from the recording plane for each class. Since most of the cells flowing in our microfluidic device are concentrated near the center of the channel by the sheath flow, we manually adjusted the z-position of the microfluidic channel to obtain holograms at varying distances from the focal plane. A distribution of particle distances from the hologram plane for each class in the yeast metabolic state dataset is presented in Fig. 3. For each metabolic state, 1000 holograms were captured at varying positions along the imaging channel at a frame rate of 12 frames per second (FPS).

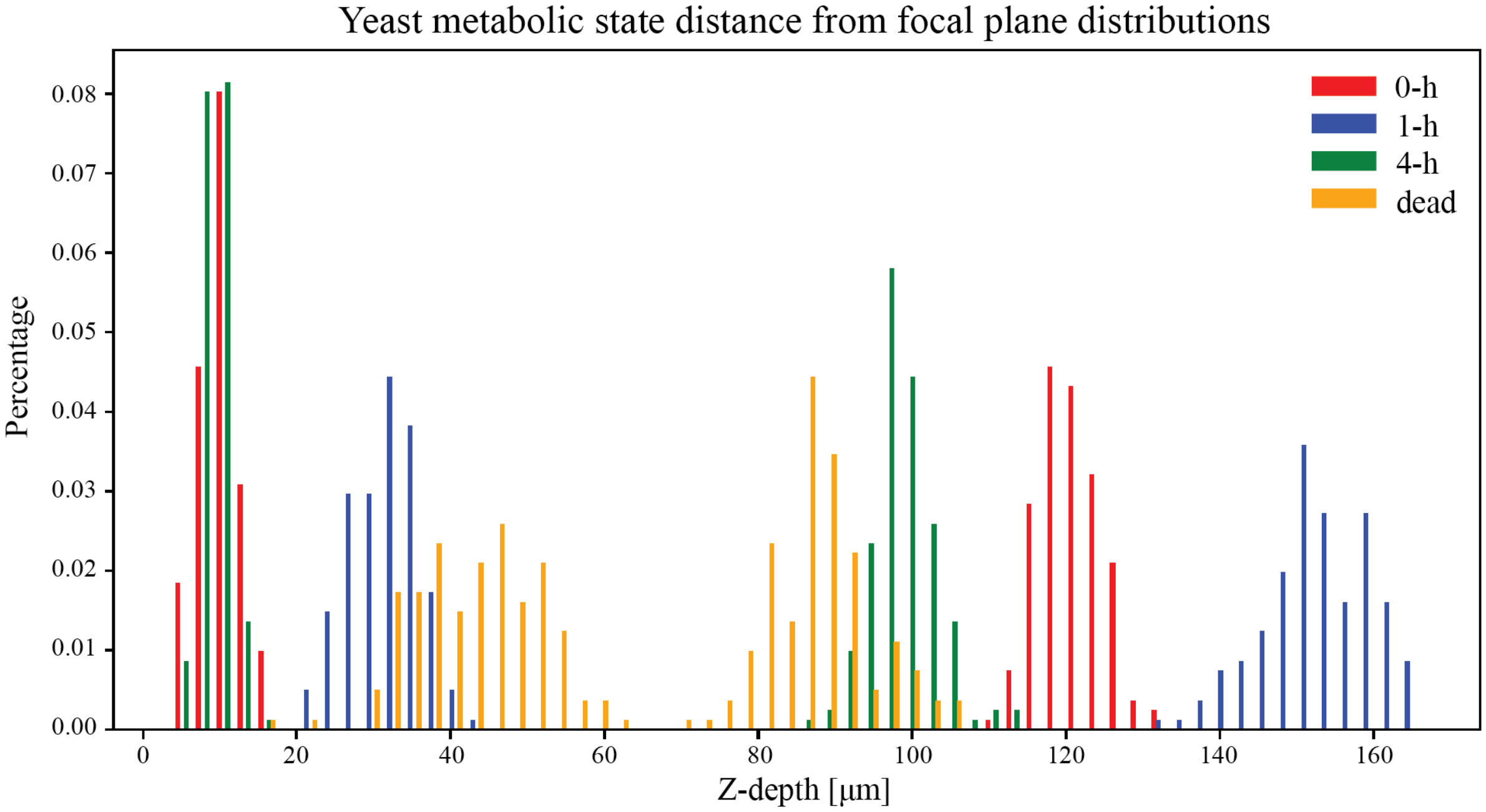

**Fig. 3.** The distribution of imaging distance from focal plane for yeast cells included in the metabolic dataset. Different colors represent cells from the four metabolic states (0-h, 1-h, 4-h, and dead) respectively. Two imaging distances were used for each class and can be identified by distribution peaks.

During the process of generating the labeled dataset for the metabolic states and strains, attached mother and daughter cells, as a result of the natural reproduction during fermentation, were often observed and labeled as a single object, while overlapping cells were given separate labels. We were able to differentiate attached mother and daughter yeast cells from overlapping cells based on the differences in the appearance of their diffraction patterns. The training set for model training consisted of both recorded and synthetic holograms. The recorded holograms from experiments only contained yeast cells from one metabolic group (or one strain), while synthetic holograms were generated by blending two or more recorded holograms randomly selected from the four metabolic groups (or two strains). This resulted in a mixture of yeast cells at various metabolic states (or from different strains) being present in the synthetic holograms. A total of 500 synthetic holograms were generated for the training set, which was then augmented by adjusting contrast, brightness, and adding random Gaussian noise, resulting in a 3-fold increase in the training size. The numbers of captured single cells contained in the training and test sets are listed in Table 2. For model performance evaluation, an additional 100 synthetic holograms were generated for the test set. For yeast strain classification, samples of lager yeast were prepared in a similar manner to the ale yeast. Lager strains are bottom fermenting and works at lower temperatures (8-15 ºC), in contrast to ale yeast which prefers higher temperatures (18-22 ºC). As shown in Fig. 2, lager yeast does not grow when cultured at 30 ºC, as indicated by the unchanged OD 600 over time.

**Table 2.** Summary of the number of yeast cells used for training and testing.

| | Ale | | | | | Lager |
|---|---|---|---|---|---|---|
| | 0h | 1h | 4h | Dead | Inactive | Inactive |
| Train | 7290 | 5572 | 9094 | 14947 | 540 | 362 |
| Test | 2163 | 1610 | 2641 | 1157 | 64 | 99 |

To assess the performance of the models for classification in each test case, we consider three criteria including precision, recall, and overall extraction rate. Precision, $P_n$, is the proportion of true positives ($TP_n$) to the total predictions made for a class $n$ and is calculated as

$$P_n = \frac{TP_n}{TP_n + FP_n} \tag{1}$$

where $FP_n$ is the number of cells which are mislabeled as class $n$, or false positives. Instances of background false positives occur when a model detects debris and abiotic particles or a small region of image background as yeast cells, but these occurrences are rare. For example, in the plankton and metabolic yeast test cases, background false positives account for only 3.9% and 1.7% of the total predictions, respectively. Therefore, background false positives are excluded from the calculation of precision when evaluating the classification performance for the convenience of direct comparison to other classifier models which do not encounter background false positives.

Recall is the percentage of the correctly classified objects in a class $n$ compared to the ground truth and is calculated as

$$R_n = \frac{TP_n}{TP_n + FN_n} \tag{2}$$

where $FN_n$ is the number of false negatives, including cells that are either mislabeled as other classes or undetected (missed cells with no bounding boxes). Since the YOLO model performs detection and classification simultaneously (one-stage), cells that do not reach the selected confidence level to be classified into any of the object classes will appear as undetected (see Fig. 6b). This is different from the conventional classification models which always assign each cell to one class.

We also use extraction rate (EA) to monitor the portion of undetected yeast cells ($N_{miss}$) as part of the comprehensive evaluation of our model. Extraction rate is calculated as

$$EA = 1 - \frac{N_{miss}}{N_{total}} \tag{3}$$

where $N_{total}$ is the total number of cells in all classes. It is common for the extraction rate to decrease if the model's classification performance declines. The confidence threshold for model prediction is selected based on the F1 score, a harmonic mean of precision and recall. Fig. 4 shows an example how the F1 score varies with different confidence thresholds. The confidence threshold corresponding to the peak of F1 score is commonly selected to ensure highest possible precision and recall rates for the model, without emphasizing one over the other.

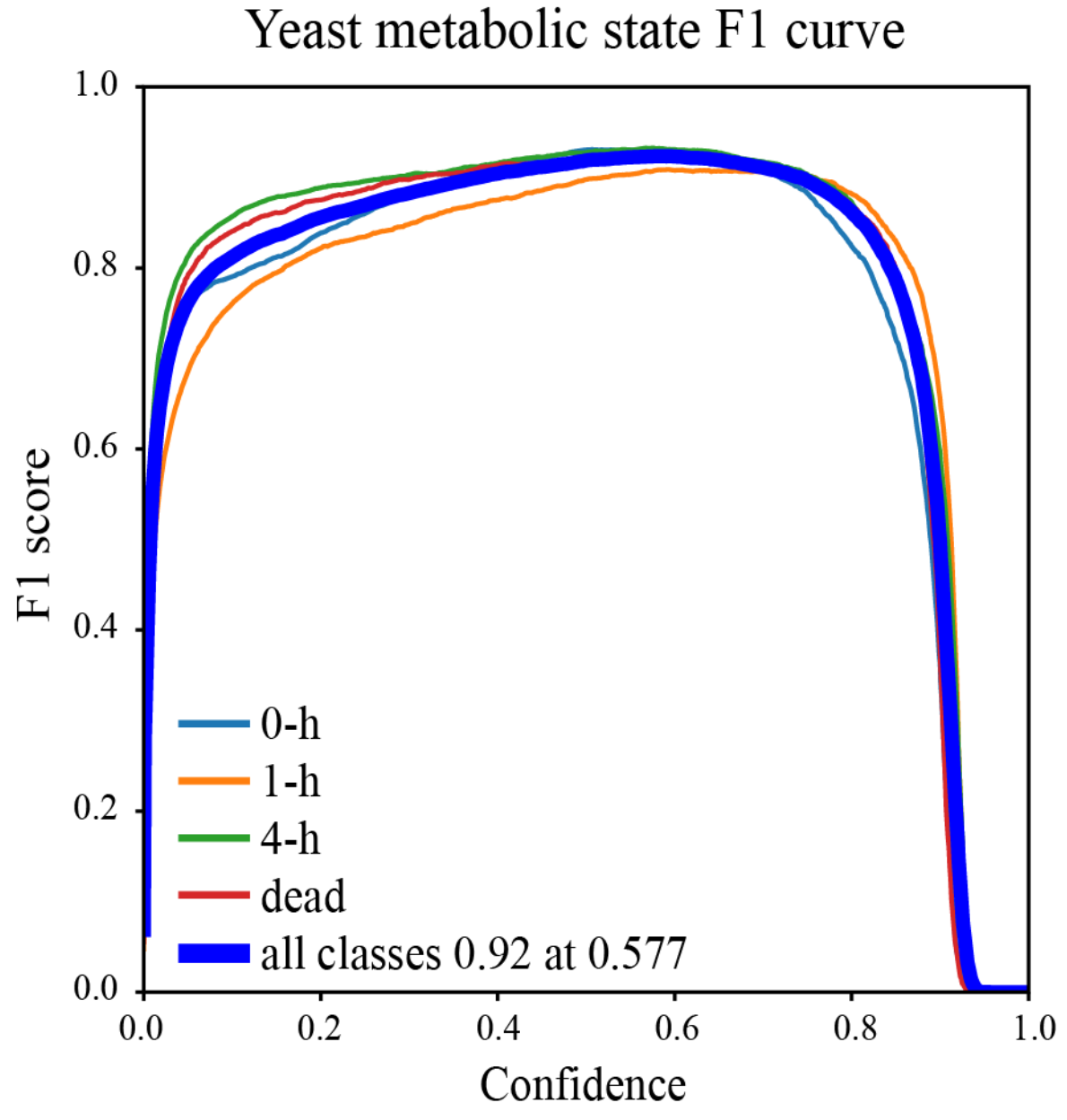

**FIG. 4.** The change of F1 score with confidence threshold for each class for the YOLO model in the yeast metabolic state case. The peak of the curve represents the highest possible precision and recall rates for the model, without emphasizing one over the other.

## III. RESULTS

Three cases were presented to demonstrate that the proposed image-based *in situ* particle analyzer is applicable to a variety of tasks in scientific research and industrial applications. The first case involved classifying plankton from 10 different species, and demonstrated an improvement in processing speed compared to the previous method (Guo et al., 2021) while maintaining a similar level of accuracy. The second case demonstrated the capability of the proposed method to differentiate yeast cells under four metabolic states from holograms without reconstruction. The third case showcased the ability of the proposed method to detect subtle differences in the subcellular structures such as the biochemical compositions of two different strains of yeast cells under the same fermentation conditions.

### A. Analysis of 10 different plankton species

Plankton are incredibly important to aquatic ecosystems and play a significant role in various research areas such as aquatic ecology, ocean optics, and climate change. Sudden uncontrolled growth of specific types of plankton species can lead to HABs, which can be detrimental to both aquatic ecosystems and human health. The timing of the initiation of HABs and its dynamics are often unpredictable due to varying environmental conditions year from year and a lack of understanding of the specific triggers and growth factors involved. Accurate detection of different plankton species is essential for long-term *in situ* monitoring of dynamic changes of their concentration which help improve our fundamental understanding of HABs. While DIH has been successfully applied for aquatic particle and HAB monitoring (Nayak et al., 2019) and automatic plankton classification (Guo et al., 2021), the computational intensity of image preprocessing has largely restricted its use in real-time analysis. Currently, it is unclear whether any neural network model is able to directly detect and distinguish different plankton species from unreconstructed holograms without any costly preprocessing. As shown in Fig. 5a, some species do not have distinctive holographic signatures (e.g., *Copepod Nauplii vs. Copepod*, and *Tripos cf. furca vs. Tripos cf. muelleri*), which may pose a challenge for accurate identification.

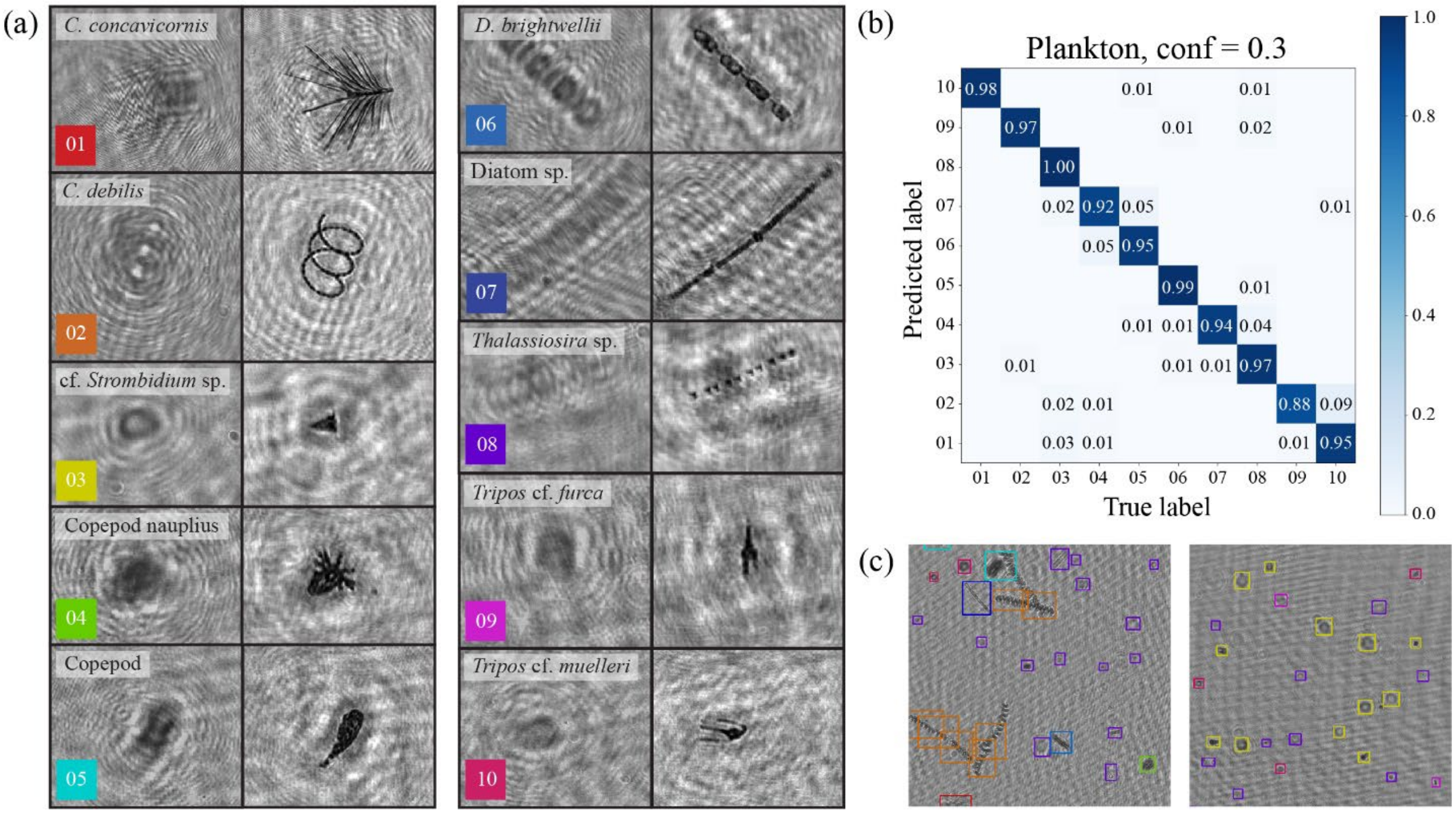

**FIG. 5.** (a) Sample holograms from each of the 10 plankton species (left) and their corresponding reconstructed in-focus images (right). (b) The confusion matrix summarizing the accuracy and prediction errors made by the model. (c) Prediction of plankton species distributed in a sample hologram with different types of planktons presented. Bounding boxes with different colors indicate different plankton species.

Our trained YOLO model was tested at a confidence threshold of 0.3 with an overall extraction rate of 99.8%. The confusion matrix in Fig. 5b summaries the accuracy and prediction errors of the model for each plankton species. Most species are classified correctly with a precision greater than 90% (dark blue boxes). The average precision is 95.3% with a 91.7% recall. The slightly lower prediction accuracy for *Tripos* cf. *furca* (88%) is likely due to its diffraction patterns being similar to some other plankton species such as *Tripos* cf. *muelleri* (9% of the predicted *Tripos* cf. *furca* cells are actually *Tripos* cf. *muelleri*). Our method achieved a similar average precision to that of the method proposed by Guo et al. (2021) (95.3% vs. 96.8%) with slightly lower recall (91.7% vs. 95.0%). The method proposed by Guo et al. (2021) involves intensive image preprocessing even without reconstruction which takes 1.6 – 2.5 seconds to process a single image. By contrast, our proposed method uses the pre-trained plankton detection model which is able to perform real-time, *in situ* analysis of captured holograms after minimal processing (enhancement), processing over 40 frames per second (a single image in 0.025 seconds). Additionally, our method was applied to holograms selected from an *in situ* dataset recorded by the HOLOCAM on 21 September 2015 at East Sound (Guo et al., 2021), which contain multiple different plankton species (Fig. 5c). Most plankton were detected successfully in these holograms, demonstrating the potential of our method as an *in situ* monitoring tool for plankton species in water.

### B. Analysis of yeast cells under different metabolic states

We used our proposed method to detect yeast cells under four different metabolic states. Based on the growth curve (Fig. 2), the holograms of ale yeast were captured at the start of fermentation (0h), during the lag phase (1-h), during the log phase (4-h), and when the cells were dead (54-h). Fig. 6a shows examples of the enhanced holograms of yeast cells from these four groups. Yeast cells from the same metabolic group may appear in different sizes and with different fringe patterns (spacing and width of fringes) depending on their distance to the focal plane during recording (small zoomed in figures below each enhanced hologram in Fig. 6a). Our dataset includes cells that could appear anywhere from 0 to 250 μm from the focal plane (Fig. 3). Despite the lack of visible differences in the holograms of single cells between groups, our trained machine learning model was able to accurately classify individual yeast cells from the four groups (colored bounding boxes in Fig. 6a). To further demonstrate the model's classification capability, synthetic holograms were generated by blending randomly selected two or more enhanced holograms from different groups (Fig. 6b). These synthetic holograms contain a mixture of yeast cells from different metabolic states. When tested on the synthetic holograms, the machine learning model was also able to accurately classify the yeast cells from the four groups.

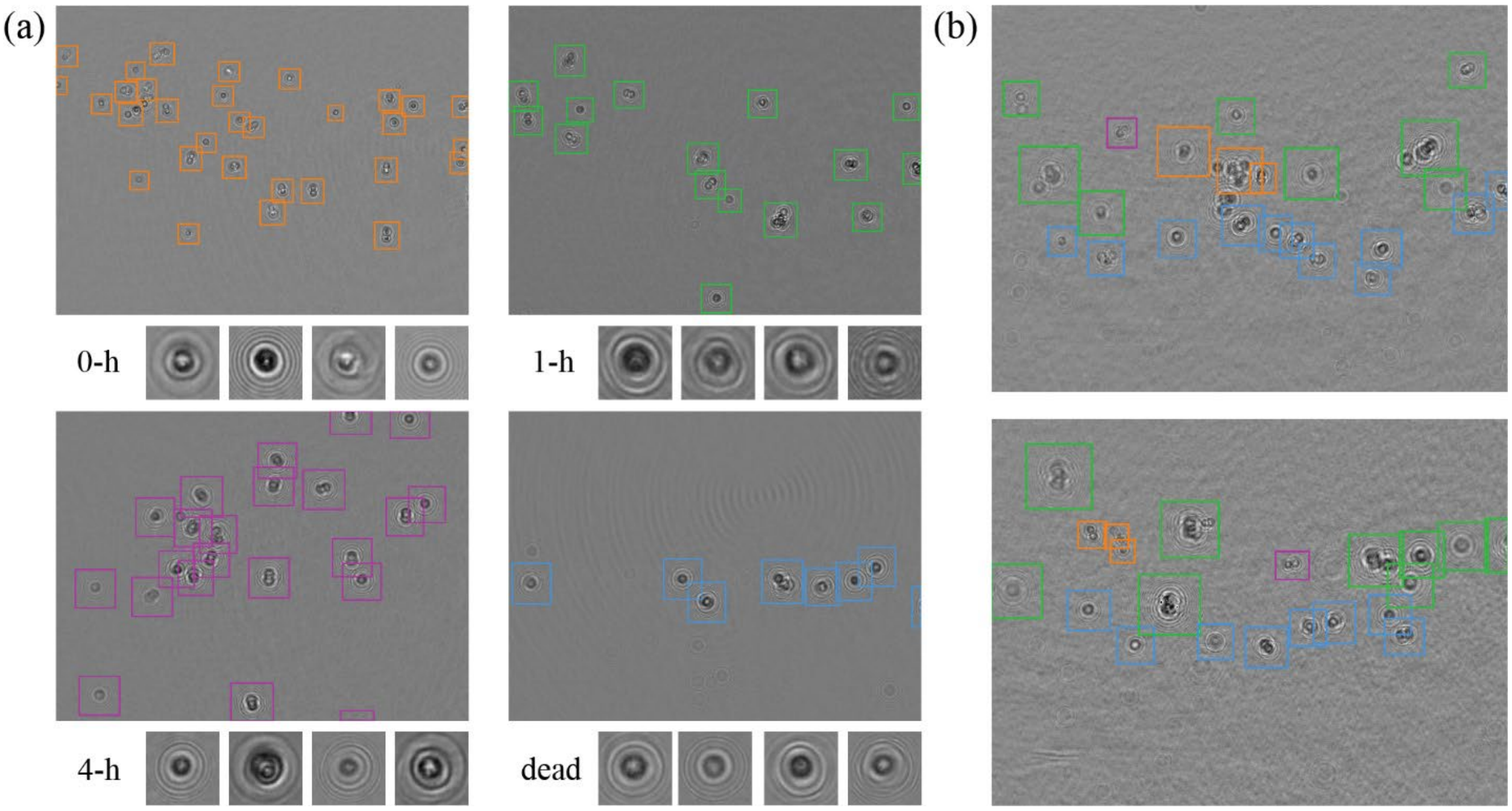

**FIG. 6.** (a) YOLO detection on the enhanced holograms of ale yeast at hour 0 (0-h), 1 (1-h), 4 (4-h) and 54 (dead) during fermentation. Samples of holograms of individual ale yeast cells are shown below. (b) YOLO detection on the synthetic holograms. Yeast cells under four different metabolic states (0-h, 1-h, 4h and dead) are predicted by bounding boxes with orange, green, purple, and blue color respectively. The numbers on the bound boxes represent the prediction confidence score.

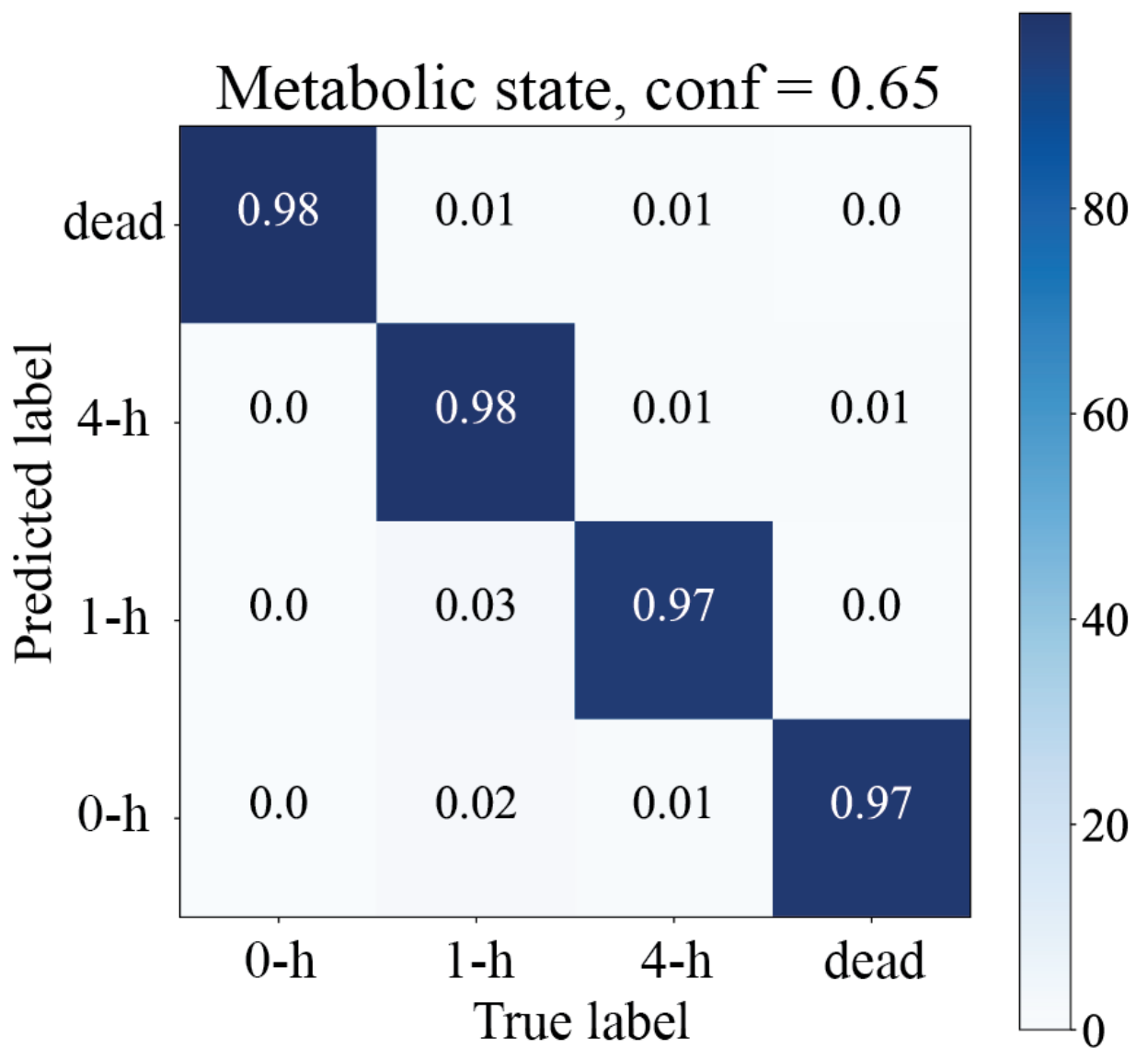

**FIG. 7.** The confusion matrix summarizing the accuracy and prediction errors made by proposed YOLO based method for classifying yeast cells under four metabolic states. The diagonal elements (dark blue boxes) indicate the precision.

We used synthetic holograms (Fig. 6b) to evaluate the model's performance in classification. A confidence threshold of 0.65 was selected according to the F1 score (Fig. 4). The accuracy and prediction errors for each group were summarized in a confusion matrix shown in Fig. 7. Each column represents the true metabolic group of each cell and each row lists the predicted metabolic groups. The diagonal elements of the matrix show the percentage of correctly classified cells from each metabolic group, or precision. For each group, the precision is greater than 96% (0-h: 98.6%, 1-h: 98%, 4-h: 96.5%, and dead: 97%). The average precision across all conditions is 97.5% with an overall extraction rate of 90.8%. The average recall is 88.5% and remains relatively constant across each group (0-h: 88%, 1-h: 87%, 4-h: 90%, and dead: 89.4%). Since we utilized a one-stage YOLO model which performs detection and classification simultaneously, the recall rate is affected by both misclassification and missed detection. This is different from the reported recall of the conventional classifier models applied after the detection, which only considers misclassification as the false negative. Our average recall is 97.3% if the missed detections are excluded, to be compared directly to the conventional classifier models. Overall, our results demonstrate the potential of *in situ* monitoring of the metabolic states of yeast cells during industrial production.

### C. Analysis of yeast cells of two different strains

Different strains of yeast cells can have different fermentation characteristics and contribute unique flavor to beer (Mochaba et al.,1998; Heggart et al., 2000). It is important in the brewing industry to maintain the purity of yeast strains in order to ensure product quality and it is therefore crucial to detect any contamination from wild yeast or undesired strains. *Saccharomyces cerevisiae* (ale yeast) and *Saccharomyces pastorianus* (lager yeast) are two common yeast strains used in beer fermentation (Bonatto 2021). Lager strains are bottom fermenting and thrive at lower temperature (8-15 ºC), while ale yeast is top fermenting and works best at room temperature (18-22 ºC). When both yeast cells were cultured at the same condition (as described in the Materials and Methods) at 30 ºC, lager yeasts remained inactive, as indicated by the unchanged cell density (Fig. 2), in contrast to the rapid growth of ale yeast. Holograms were captured of cells samples 1 hour after the initiation of fermentation for both ale and lager groups (Fig. 8a). The image resolution and variation in diffraction pattern from varying imaging depths are the same as the metabolic state experiments. Synthetic holograms containing a mixture of cells from both groups were also generated using blending (Fig. 8b). The trained YOLO model was able to accurately distinguish ale yeast from lager yeast in both recorded (Fig. 8a) and synthesized (Fig. 8b) holograms, despite their similar appearances (zoomed in images in Fig. 8a). The precision is 99.5% for ale yeast and 97% for lager yeast, with an average of 98.2% (Fig. 9). The recall for the ale yeast is 97.8% and 83.1% for the lager, resulting in an average of 90.5%. The overall extraction rate is 96.4%. The slightly lower recall for the lager yeast may be due to the presence of a higher number of abiotic particles in the recorded holograms, which could interfere with the yeast cells and hinder the model's ability to extract features.

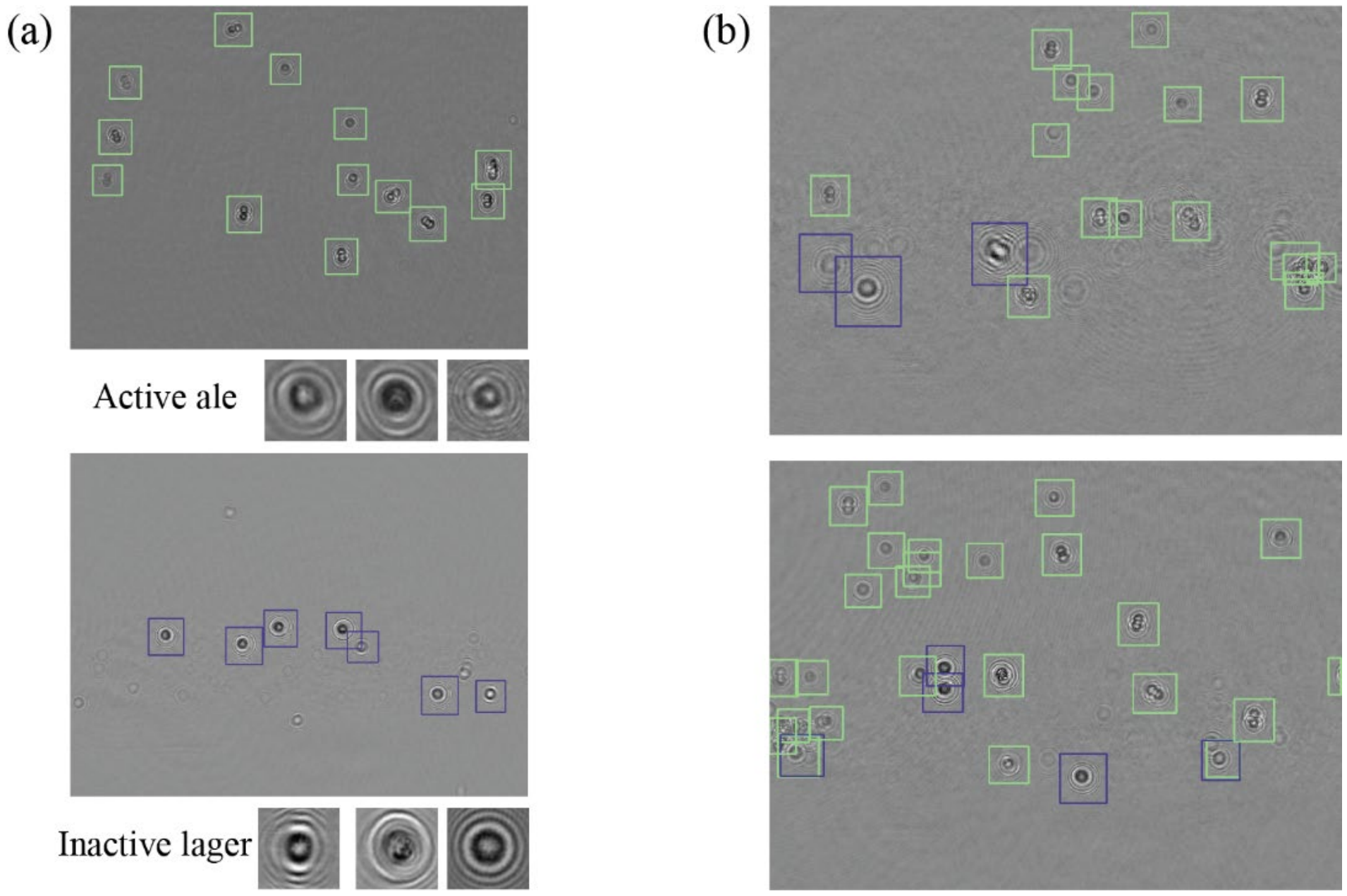

**FIG. 8.** (a) YOLO detection on the enhanced holograms of ale yeast and lager yeast at hour 1 during fermentation. Sample holograms of individual ale and lager yeast cells are shown below. (b) Detection result showcases for the synthetic holograms. Ale yeast cells are predicted by the purple bounding boxes and the lager yeast cells are predicted by the beige bounding boxes. The numbers on the bound boxes represent the confidence score.

As a control experiment, we also captured holograms of inactive ale yeast cells, which were prepared by directly dissolving the dry ale yeast in distilled water without culturing. The average precision and recall were only 62% and 55.6%, respectively, indicating that the YOLO model is not able to distinguish between ale and lager yeast when both are inactive. The overall extraction rate is 87.2%. The lower extraction rate is likely due to the model’s poor performance in classification. Because the YOLO model performs detection and classification simultaneously, more cells will be undetected if the model fails to classify them correctly. If the model is only trained on a detection task for any yeast cells, the extraction rate is 93%, comparable to the other yeast test cases. These results suggest that our method can distinguish different strains based on their unique metabolic characteristics during fermentation.

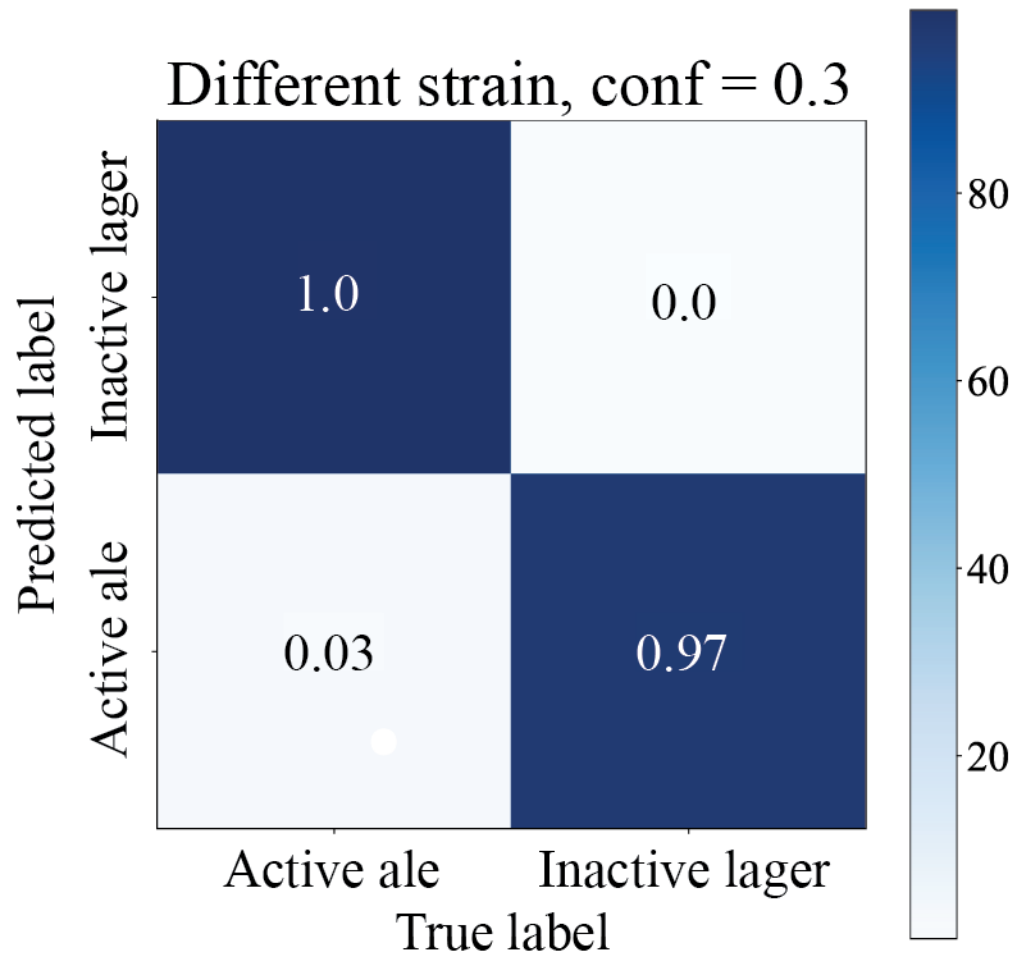

**FIG. 9.** The confusion matrix summarizing the accuracy and prediction errors made by proposed YOLO based method for yeast strain detection during fermentation. The diagonal elements (dark blue boxes) indicate precision of the model prediction.

## IV. CONCLUSIONS AND DISCUSSIONS

In this study, we introduce a novel approach for real-time *in situ* analysis of biological particles using machine learning assisted digital inline holography (DIH). Our machine learning model, which uses a modified YOLO v5 architecture customized for the detection and classification of holograms of small biological particles, is optimized using TensorRT for real-time processing. Unlike previous methods used to classify particles captured in holograms (Kim et al., 2018; Guo et al., 2021), our approach integrates particle localization and classification into a single step, significantly reducing processing time while maintaining prediction accuracy. We have demonstrated the capability of our novel approach for *in situ* biological particle analysis using three test cases: classifying 10 different species of plankton, detecting yeast cells under four different metabolic states, and differentiating two yeast strains during fermentation. Our approach does not require any additional preprocessing (e.g., hologram reconstruction and particle segmentation) used in other studies (Kim et al., 2018; Guo et al., 2021), significantly reducing processing time and computational resources. Our DIH sensors, with onboard processing capability, are ideal for real-time, *in situ* monitoring of the onset and development of harmful algal blooms, or the viability and vitality of yeast cells during various industrial processes. Our method is sensitive enough to detect subtle biochemical composition changes in single cells (Raschke & Knorr, 2009; Chan et al., 2012) and can also be used to distinguish different strains of yeast cells based on their unique fermentation characteristics.

Overall, this work showcases the potential of machine learning assisted DIH as a novel and versatile tool for real-time, *in situ* analysis of various biological particles, including their morphology, viability, vitality, and other important biophysical properties that are correlated with changes in optical density (e.g., membrane structure, protein). Compared to commonly used rapid particle analysis tools such as laser diffraction, acoustic scattering, and Coulter counter, our approach offers more information beyond size and concentration. It can be used for viability and vitality tests similar to conventional fluorescent microscopy and flow cytometry methods, but without the need for sample preparation (label-free) and with significantly higher throughput. Our optical setup is highly compact and cost-effective compared to other label-free technologies such as QPI and hyperspectral imaging, and the data process is not as computationally intensive with much higher throughput. With these features, our proposed method can be easily extended to the analysis of other types of particles (both biotic and abiotic) and can be deployed in a distributive manner for scientific research and manufacturing on an industrial scale. To apply the method to a specific application, holograms can be captured using similar hardware setup using customized objectives and camera sensors based on particles size and field of view. The biotic and abiotic particles in the holograms can then be manually labeled by a domain expert to form a training set and a machine learning model using our proposed architecture can be trained. Once the model is trained and loaded onboard, the entire system should be able to perform effectively for the application.

However, it is important to note that the precision of our machine learning model may decrease when analyzing holograms with high particle concentrations due to the overlap of diffraction patterns, but this will not significantly influence the model's performance as long as the overlap is less than 50%. This conclusion is based on test results that showed high precision and recall in most cases. Our method is particularly valuable for analyzing particles in the applications that require high sensitivity, such as biocontaminants in sterile liquids (e.g., spring water, sterile liquid used in pharmaceutical industries and clinical applications). In these applications, particle concentrations are usually low and our model performance will not be significantly affected by overlapping fringes. In addition, despite the generalizability of our overall approach including the hardware setup and ML method, the trained ML model is specific to the application it is trained for, and its performance may be compromised when used on data outside the scope of its training. This is the limitation for common supervised learning approaches. To ensure the accuracy and robustness of the model, it may be necessary to retrain it and gather new labeled data when there are changes in particle properties (e.g., size, shape), medium properties (e.g., refractive index), or image acquisition settings (e.g., magnification, laser wavelength). In the future, it may be possible to mitigate this limitation by using unsupervised or semi-supervised machine learning model architectures in DIH data processing.

## REFERENCES

Baddour, R. E., Sherar, M. D., Hunt, J. W., Czarnota, G. J., & Kolios, M. C. (2005). High-frequency ultrasound scattering from microspheres and single cells. *The Journal of the Acoustical Society of America*, 117(2), 934-943.

Beuthan, J., Minet, O., Helfmann, J., Herrig, M., & Müller, G. (1996). The spatial variation of the refractive index in biological cells. *Physics in Medicine & Biology*, 41(3), 369.

Bista, R. K., Uttam, S., Wang, P., Staton, K. D., Brand, R. E., Liu, Y., ... & Hartman, D. J. (2011). Quantification of nanoscale nuclear refractive index changes during the cell cycle. *Journal of Biomedical Optics*, 16(7), 070503.

Bonatto, D. (2021). The diversity of commercially available ale and lager yeast strains and the impact of brewer's preferential yeast choice on the fermentative beer profiles. *Food Research International*, 141, 110125.

Chan, L. L., Kury, A., Wilkinson, A., Berkes, C., & Pirani, A. (2012). Novel image cytometric method for detection of physiological and metabolic changes in Saccharomyces cerevisiae. *Journal of Industrial Microbiology and Biotechnology*, *39*(11), 1615-1623.

Choi, S. Y., Oh, J., Jung, J., Park, Y., & Lee, S. Y. (2021). Three-dimensional label-free visualization and quantification of polyhydroxyalkanoates in individual bacterial cell in its native state. *Proceedings of the National Academy of Sciences*, 118(31), e2103956118.

Choi, W. J., Jeon, D. I., Ahn, S. G., Yoon, J. H., Kim, S., & Lee, B. H. (2010). Full-field optical coherence microscopy for identifying live cancer cells by quantitative measurement of refractive index distribution. *Optics Express*, 18(22), 23285-23295.

Coling, D., & Kachar, B. (1998). Principles and application of fluorescence microscopy. *Current Protocols in Molecular Biology*, 44(1), 14-10.

Croce, A. C., & Bottiroli, G. (2014). Autofluorescence spectroscopy and imaging: a tool for biomedical research and diagnosis. *European Journal of Histochemistry*, 58(4), 2461.

Davey, H. M., & Kell, D. B. (1996). Flow cytometry and cell sorting of heterogeneous microbial populations: the importance of single-cell analyses. *Microbiological Reviews*, 60(4), 641-696.

Eady, M., Park, B., & Choi, S. U. N. (2015). Rapid and early detection of Salmonella serotypes with hyperspectral microscopy and multivariate data analysis. *Journal of Food Protection*, 78(4), 668-674.

Glenn Jocher, Alex Stoken, Ayush Chaurasia, Jirka Borovec, NanoCode012, TaoXie, Yonghye Kwon, Kalen Michael, Liu Changyu, Jiacong Fang, Abhiram V, Laughing, tkianai, yxNONG, Piotr Skalski, Adam Hogan, Jebastin Nadar, imyhxy, Lorenzo Mammana, …. (2021). ultralytics/yolov5: v6.0 - YOLOv5n 'Nano' models, Roboflow integration, TensorFlow export, OpenCV DNN support (v6.0). Zenodo. https://doi.org/10.5281/zenodo.5563715

Guo, B., Nyman, L., Nayak, A. R., Milmore, D., McFarland, M., Twardowski, M. S., ... & Hong, J. (2021). Automated plankton classification from holographic imagery with deep convolutional neural networks. *Limnology and Oceanography: Methods*, 19(1), 21-36.

Heggart, H., Margaritis, A., Stewart, R. J., Pilkington, M., Sobezak, J., & Russell, I. (2000). Measurement of brewing yeast viability and vitality: a review of methods. *Technical quarterly-Master Brewers Association of the Americas*, 37(4), 409-430.

Ho, C. S., Jean, N., Hogan, C. A., Blackmon, L., Jeffrey, S. S., Holodniy, M., ... & Dionne, J. (2019). Rapid identification of pathogenic bacteria using Raman spectroscopy and deep learning. *Nature Communications*, 10(1), 1-8.

Ho, J. C., & Michalak, A. M. (2015). Challenges in tracking harmful algal blooms: A synthesis of evidence from Lake Erie. *Journal of Great Lakes Research*, 41(2), 317-325.

Kaikkonen, V. A., Ekimov, D., & Mäkynen, A. J. (2014). A holographic in-line imaging system for meteorological applications. *IEEE Transactions on Instrumentation and Measurement*, 63(5), 1137-1144.

Kang, R., Park, B., Eady, M., Ouyang, Q., & Chen, K. (2020). Single-cell classification of foodborne pathogens using hyperspectral microscope imaging coupled with deep learning frameworks. *Sensors and Actuators B: Chemical*, 309, 127789.

Katz, J., & Sheng, J. (2010). Applications of holography in fluid mechanics and particle dynamics. *Annual Review of Fluid Mechanics*, 42(1), 531-555.

Kim, S. J., Wang, C., Zhao, B., Im, H., Min, J., Choi, H. J., ... & Lee, K. (2018). Deep transfer learning-based hologram classification for molecular diagnostics. *Scientific Reports*, 8(1), 1-12.

Kumar, S., Zou, S., Sun, Y. & Hong, J. (2016). 3D holographic observatory for long-term monitoring complex behaviors in drosophila. *Scientific Reports* 6. doi.org/10.1038/srep33001

Li, J., Shao, S. & Hong, J. (2020). Machine learning shadowgraph for particle size and shape characterization. *Measurement Science & Technology*, 32, 015406.

Nayak, A. R., McFarland, M. N., Twardowski, M. S., Sullivan, J. M., Moore, T. S., & Dalgleish, F. R. (2019, May). Using digital holography to characterize thin layers and harmful algal blooms in aquatic environments. In *Digital Holography and Three-Dimensional Imaging* (pp. Th4A-4). Optical Society of America.

Neubeck, A., & Van Gool, L. (2006, August). Efficient non-maximum suppression. In *18th International Conference on Pattern Recognition (ICPR'06)* (Vol. 3, pp. 850-855). IEEE.

Nguyen, N. D., Do, T., Ngo, T. D., & Le, D. D. (2020). An evaluation of deep learning methods for small object detection. *Journal of Electrical and Computer Engineering*, 3189691

Nyquist, H. (1928). Certain topics in telegraph transmission theory. *Transactions of the American Institute of Electrical Engineers*, 47(2), 617-644.

Maas, A. L., Hannun, A. Y., & Ng, A. Y. (2013, June). Rectifier nonlinearities improve neural network acoustic models. In *Proc. icml* (Vol. 30, No. 1, p. 3).

Mallery, K. & Hong, J. (2019). Regularized inverse holographic volume reconstruction for PTV. *Optics Express*, 27(13), 18069-18084.

Maltsev, V. P., & Semyanov, K. A. (2013). Characterisation of bio-particles from light scattering. In *Characterisation of Bio-Particles from Light Scattering*. De Gruyter.

Mochaba, F., O'Connor-Cox, E. S. C., & Axcell, B. C. (1998). Practical procedures to measure yeast viability and vitality prior to pitching. *Journal of the American Society of Brewing Chemists*, 56(1), 1-6.

Popescu, G. (2011). Quantitative phase imaging of cells and tissues. McGraw-Hill Education.

Raschke, D., & Knorr, D. (2009). Rapid monitoring of cell size, vitality and lipid droplet development in the oleaginous yeast Waltomyces lipofer. *Journal of microbiological methods*, 79(2), 178-183.

Redmon, J., Divvala, S., Girshick, R., & Farhadi, A. (2016). You only look once: Unified, real-time object detection. In *Proceedings of the IEEE conference on computer vision and pattern recognition* (pp. 779-788).

Sauvageat, E., Zeder, Y., Auderset, K., Calpini, B., Clot, B., Crouzy, B., ... & Vasilatou, K. (2020). Real-time pollen monitoring using digital holography. *Atmospheric Measurement Techniques*, 13(3), 1539-1550.

Shao, S., Mallery, K., Kumar, S. S., & Hong, J. (2020a). Machine learning holography for 3D particle field imaging. *Optics Express*, 28(3), 2987-2999.

Shao, S., Mallery, K., & Hong, J. (2020b). Machine learning holography for measuring 3D particle distribution. *Chemical Engineering Science*, 225, 115830.

Shapiro, H. M. (2004). The evolution of cytometers. C*ytometry Part A: The Journal of the International Society for Analytical Cytology*, 58(1), 13-20.

Stephens, D. J., & Allan, V. J. (2003). Light microscopy techniques for live cell imaging. *Science*, 300(5616), 82-86.

Strola, S. A., Baritaux, J. C., Schultz, E., Simon, A. C., Allier, C., Espagnon, I., Jary, D. & Dinten, J. M. (2014). Single bacteria identification by Raman spectroscopy. *Journal of Biomedical Optics*, 19(11), 111610.

Sun, T., & Morgan, H. (2010). Single-cell microfluidic impedance cytometry: a review. *Microfluidics and Nanofluidics*, *8*(4), 423-443.

Sung, Y., Choi, W., Lue, N., Dasari, R. R., & Yaqoob, Z. (2012). Stain-free quantification of chromosomes in live cells using regularized tomographic phase microscopy. *PloS One*, 7(11), e49502.

Yan, J., Lei, Z., Wen, L., & Li, S. Z. (2014). The fastest deformable part model for object detection. In *Proceedings of the IEEE Conference on Computer Vision and Pattern Recognition* (pp. 2497-2504).

Yoon, S. C., Lawrence, K. C., Siragusa, G. R., Line, J. E., Park, B., & Feldner, P. W. (2009). Hyperspectral reflectance imaging for detecting a foodborne pathogen: Campylobacter. *Transactions of the ASABE*, 52(2), 651-662.

You, J., Mallery, K., Mashek, D., Sanders, M., Hong, J. & Hondzo, M. (2020). Microalgal swimming signatures as indicators of neutral lipids production across growth phases. *Biotechnology and Bioengineering*, 117(4), 970-980.

You, J., Mallery, K., Hong, J. & Hondzo, M. (2018) Temperature effects on growth and migration of Microcystis. *Journal of Plankton Research*, 40(1), 16-28.

Yu, Z. T. F., Aw Yong, K. M., & Fu, J. (2014). Microfluidic blood cell sorting: now and beyond. *Small*, 10(9), 1687-1703.